\documentclass{svproc}
\usepackage{url}
\usepackage{graphicx}
\usepackage{amsfonts,amssymb,amsmath,mathrsfs}
\usepackage{color}
\usepackage{braket}

\begin{document}
\mainmatter           
\title{Jack on a Devil's staircase}  
\titlerunning{Jack on a Devil's staircase}  
\author{ Andrea Di Gioacchino\inst{1}, Marco Gherardi\inst{1}, Luca Guido Molinari\inst{1}\\ and Pietro Rotondo\inst{2,3}}
\authorrunning{A. Di Gioacchino, M. Gherardi, L. G. Molinari, P. Rotondo}
\institute{Dipartimento di Fisica, Universit\`a degli Studi di Milano, and I.N.F.N. sezione di Milano,
Via Celoria 16, 20133 Milano, Italy
\and
School of Physics and Astronomy, University of Nottingham, Nottingham, NG7 2RD, UK
\and
Centre for the Mathematics and Theoretical Physics of Quantum Non-equilibrium Systems,
University of Nottingham, Nottingham NG7 2RD, UK
}
\maketitle       

\begin{abstract}
We review a simple mechanism for the formation of plateaux in
the fractional quantum Hall effect. It arises from a map of the
microscopic Hamiltonian in the thin torus limit 
to a lattice gas model, solved by Hubbard. The map suggests
a Devil's staircase pattern, and explains the observed asymmetries
in the widths. Each plateau is a new ground state of the system:
a periodic Slater state in the thin torus limit.
We provide the unitary operator that maps such limit states
to the full, effective ground states with same filling fraction. These 
Jack polynomials generalise Laughlin's ansatz, and are exact 
eigenstates of the Laplace-Beltrami operator. Why are Jacks 
sitting on the Devil's staircase? This is yet an intriguing problem.
\keywords{Quantum Hall effect, Laughlin ansatz, Jack polynomials, Laplace Beltrami operator}
\end{abstract}
\section{\bf The quantum Hall effects}
\begin{figure}[h]
\begin{center}
\includegraphics[width=7.5cm]{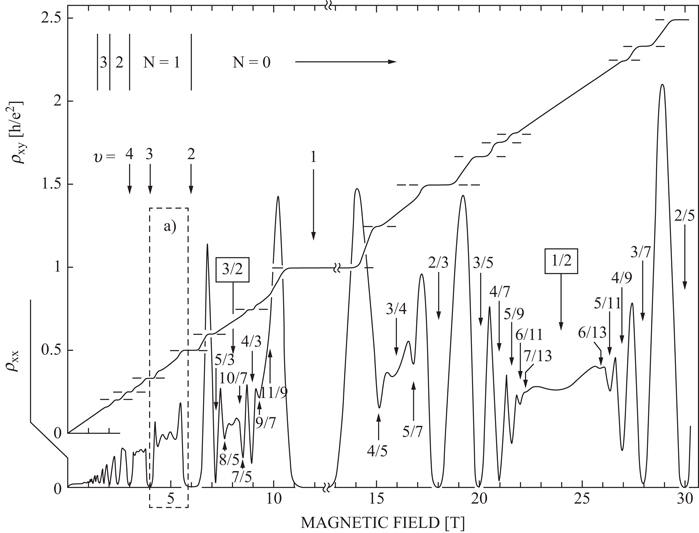}
\caption{The experimental ladder of plateaux for $\rho_{\rm xy}$ (in units of ${\rm R_H}$) as a function
of the magnetic field. Note the asymmetry in width of the plateaux $\nu $ and $1-\nu $.
}\label{figure:QHE}
\end{center}
\end{figure}

Since the discoveries of the integer (von Klitzing, 1980) and of the fractional (Tsui and Stormer, 1981) 
quantum Hall effects, the phenomena have been an amazing source of inspiration
for experimental and theoretical physics, with deep intersections with mathematics. After decades, the experimental results stand neat and beautiful (fig.\ref{figure:QHE}). \\
A gas of high mobility electrons at the interface of an heterostructure, refrigerated at mK temperature and in magnetic fields of one or several tesla, exhibits a quantisation 
of the resistance orthogonal to the current, in units of a fundamental constant,
at integer or fractional values of $\nu$: 
\begin{equation}
{\rm R_{xy} }= \left(\frac{h}{e^2} \right) \frac{1}{\nu}
\end{equation}
The parameter is the filling fraction $\nu =n/g$, where $n$ and $g=eB/hc$ are the number of electrons and the degeneracy of a Landau level per unit area. 
Substitution gives ${\rm R_{xy} }= B/(enc)$: this is Hall's law resulting
from the balance of the Lorentz force with the electric force created by a charge gradient. However, in a quantum regime, the transverse resistance does not vary continuously with the field $B$: the linear slope is a staircase. \\
The integer plateaux $\nu =1,2,3,\dots $ (IQHE) came first; next, by achieving higher magnetic fields, mobility, and lower temperatures, the prominent plateaux $\nu=1/3$ and $2/3$ were discovered. In the years, 60 or more fractional values have been observed (FQHE), the values $\nu<1$ corresponding to a partial filling of the lowest Landau level. 
While experiments are performed on micrometric ``Hall bars'', the theory is staged in geometries such as torus, sphere or plane, that avoid the complication of confining potentials. In a strong field, a boundary produces an edge current in a basically 1D layer whose properties differ from the bulk. Experiments on Hall bars with a close 
contact of edges, measured fluctuations of the tunnelling current (quantum shot noise)  
of fractional charge carriers  \cite{dePicciotto}.

Integer and fractional Hall effects are currently explained in two different ways.
The former is caused by the presence of impurities, which remove the degeneracy and broaden Landau levels 
into bands of localised states centred at the Landau energies, which remain conducting. As electrons are added, no variation of the current is seen (the plateau), 
until the Fermi energy crosses the next conducting Landau energy, and the current has a jump.\\  
Fractional plateaux are explained as an effect of the Coulomb interaction, but a final theory has not yet been given.
The Nobel prize winner Laughlin \cite{Laughlin} proposed the following ground states (symmetric gauge) at fillings $\nu =1/q$ :
\begin{equation}
\Psi_{1/q} (z_1,\dots z_N) = \exp \left( -\sum_{i=1}^N \frac{|z_i|^2}{2\ell^2} \right) \prod_{j<i} (z_i-z_j)^q
\end{equation}
The state is not normalised, $z_i=x_i-iy_i$, $\ell $ is the magnetic length ($\ell^2 = eB/hc$), $q$ is odd for fermions. 
It is an eigenstate of the angular momentum and it is zero when particles collide, thus keeping the Coulomb energy at bay. 
Although not exact, numerical diagonalization with small $N$ gives an impressive overlap with the true ground state. 
Haldane, Halperin and MacDonald obtained a hierarchy of ground states by adding quasi-holes to Laughlin states $\Psi_{1/q}$,  obtaining the filling
fractions $\nu = 2p/(2pq\pm 1)$. The process can be repeated, and in principle any value of filling fraction can be attained.\\
Another widely accepted scenario is the formation of ``composite fermions'', i.e. electrons dressed with $2p$ units of magnetic flux. 
The model was proposed by Jain, and extended by Halperin, Lee and Read \cite{JainBook}.
In a field $B$, the effective magnetic field felt by composites with electron density $n$ is $B^*=B-2pn(hc/e)$. The effective filling fraction 
is $\nu^* =n (hc/eB^*)$. For integer $\nu^*$ the IQHE of composite fermions occurs, corresponding to filling fractions $\nu = \nu^*/(2p\nu^* \pm 1)$.

In the Landau gauge, the Hilbert space spanned by the eigenstates of the lowest LL, consists of functions $\exp (-\sum_{i=1}^N \frac{|z_i|^2}{2\ell^2}) f(z_1,\dots,z_N)$. The exponential factor is included in the measure of a Bargmann space of analytic functions $f$. With this separation, 
a Laughlin state $\psi_{1/q}$ is the power of a Vandermonde determinant $\Delta (z) =\prod_{j>k}(z_j-z_k) $. 
This and several other functions,  such as Read-Moore and Read-Rezayi, 
are exact eigenstates of the Laplace-Beltrami operator, for the excited states of the Calogero-Sutherland Hamiltonian (an exactly solvable 1D many-particle model). 
They share the property of  factoring into a Vandermonde determinant and a symmetric Jack polynomial. This intriguing link with FQHE 
was uncovered in 2008 by Haldane and Bernevig \cite{Bernevig}.

Another route was attempted by Tao and Thouless in 1983 \cite{Tao1983}, to explain the phase diagram as an effect of a Wigner crystal arrangement of electrons.
However this possibility was ruled out because it predicts long-range solid order which is not observed in experiments. At the same time, the success of Laughlin's 
ansatz (for instance in giving a justification for the odd-denominator rule) discouraged further investigations in this direction.\\
The approach was revived in 2008 by Bergholtz and Karlhede \cite{Bergholtz}, who showed that a crystal structure
may arise in the strongly anisotropic quasi-one dimensional limit known as ``thin torus''€ limit. The result was obtained by mapping the microscopic model in the thin torus limit 
to a 1D lattice gas model, that was analytically solved by Hubbard \cite{Hubbard1978}
and by Bak and Bruinsma \cite{BakBruinsma}.
The lattice gas has a fractal phase diagram, with plateaux of the density as a function of the
chemical potential, forming a ``Devil's staircase''. Its implications for the FQHE were investigated by
Rotondo et al. \cite{Rotondo} and are in qualitative accord with the experimental diagram.
To our knowledge, it is the first attempt to produce a realistic phase diagram.

Here, we review the thin torus limit and the phase diagram discussed in ref. \cite{Rotondo}. We then
set the stage for the next result, with an introduction to the Calogero-Sutherland model and the
Laplace-Beltrami operator, whose eigenfunctions are Jack polynomials. \\
There is consensus that the ground states of FQHE are adiabatically connected to the thin torus eigenstates. This is made particularly evident in the work,  ref. \cite{DiGioacchino}, where a unitary operator is obtained that maps Slater determinants (or permanents) of a simpler diagonal theory, to the fully interacting eigenstates, such as the Laughlin states and, more generally, Jack-like states. The unitary operator, in essence, is the Dyson T-exp of the non-diagonal two-particle interaction in the Laplace-Beltrami operator.

\section{Thin torus and the Devil's staircase}
The Hamiltonian for the FQHE in the lowest Landau level is:
\begin{equation}
H =  \frac{\hbar\omega_c}{2}\sum_s a^\dagger_s a_s + \frac{1}{2} \sum_{\bf s} V({\bf s}) a^\dagger_{s_1} a^\dagger_{s_2} a_{s_4} a_{s_3}
\end{equation}
the operators $a^\dagger_s$ and $a_s$ create and destroy an electron in eigenstates $s=1\dots g$ spanning the lowest Landau level. 
$V({\bf s})=\langle s_1,s_2|v|s_3,s_4\rangle $ is the Coulomb matrix element. A background of positive charges cancels 
the Hartree term $s_1=s_3$, $s_2=s_4$. Note that the kinetic energy per particle $\hbar\omega_c/2 $ can be identified with a chemical potential,
that depends on $B$.\\
The earliest numerical calculations, for various filling fractions, were done by Yoshioka et al. \cite{Yoshioka1983,Yoshioka1984}. They considered the geometry of a periodic array of 
rectangles (torus) with commensurate area: $L_xL_y=2\pi \ell^2 g$, where $g$ is a large natural number. With the $g$ degenerate eigenstates of the
lowest LL in a rectangle, one constructs a basis of quasi-periodic eigenstates (Jacobi theta functions):
$$\theta_s (x,y) = \sum_{m\in {\rm Z}} \frac{1}{\sqrt {\ell L_y\sqrt\pi}} 
\exp \left[ - i \frac{2\pi}{L_y}(s+mg) y -\frac{1}{2\ell^2} \left (x+mL_x-\frac{2\pi}{L_y} s\ell^2 \right )^2\right ]
$$
with $0\le s\le  g-1$; the normalisation is such that $ \int_R dxdy \; \overline{\theta_s (x,y)}\theta_{s'} (x,y) = \delta_{ss'} $.\\
An electron in the rectangle $R$ interacts with the electrons
in $R$ as well as with their copies. The Coulomb interaction  
is a periodic function of the lattice:
$$ v({\bf r}) = 
\sum_{\bf m\in\mathbb Z^2} \frac{e^2}{\sqrt{ (x +m_xL_x)^2+(y+m_y L_y)^2}} 
= \frac{1}{L_xL_y}
\sum_{\bf q} \frac{2\pi e^2}{|{\bf q}|}   e^{ i {\bf q}\cdot {\bf r} } $$
where $q_x=\frac{2\pi}{L_x}n_x$ and $q_y=\frac{2\pi}{L_y}n_y$.
The Coulomb matrix elements $V({\bf s})$ are:
\begin{eqnarray}
V({\bf s}) &&=  \int_{R^2} d{\bf r}_1 d{\bf r}_2 \,
\overline{\theta_{s_1}({\bf r}_1)}\, \overline{\theta_{s_2}({\bf r}_2)} v({\bf r}_1-{\bf r}_2)
\theta_{ s_3}({\bf r}_1)\, \theta_{ s_4}({\bf r}_2) \nonumber \\
&&=\frac{1}{L_xL_y}\sum_{\bf q}   v({\bf q})  
I_{s_1, s_3}({\bf q}) I_{s_2, s_4}({-\bf q}) \nonumber
\end{eqnarray}
where $I_{ s,s'}({\bf q})=\int_R d{\bf r}
\overline{\theta_s ({\bf r})}\theta_{s'}({\bf r}) \exp(i{\bf q\cdot r}) $. 
The exact formula (2.9) in Yoshioka's paper \cite{Yoshioka1984} is obtained:
\begin{equation}
V({\bf s})
= \frac{\delta_{s_1+s_2,s_3+s_4}}{L_xL_y}\sum_{\bf q} \frac{2\pi e^2}{|{\bf q}|} e^{-\frac{\ell^2}{2}|{\bf q}|^2+iq_x\frac{2\pi\ell^2}{L_y}(s_3-s_2)}  \delta'(s_1-s_3+n_y)  \label{Yoshioka}
\end{equation}
The periodic (with period $g$) delta function $\delta' $ is used to sum on $q_y$:
$$
= 2\pi e^2\frac{\delta_{s_1+s_2,s_3+s_4}}{L_xL_y}\sum_{q_x}  e^{-\frac{\ell^2}{2}q_x^2+iq_x\frac{2\pi\ell^2}{L_y}(s_3-s_2)}
\sum_{m=-\infty}^\infty \frac{e^{-\frac{\ell^2}{2}\frac{4\pi^2}{L_y^2}(s_3-s_1+mg)^2 }}{\sqrt{ q_x^2 + \frac{4\pi^2}{L_y^2}(s_3-s_1+mg)^2 }}
$$
Now, approximate $\sum_{q_x} \approx \frac{L_x}{2\pi}\int dq_x $ and 
neglect terms $m\neq 0$ because of the exp factor. Eq.~(3) in the Tao and Thouless paper, \cite{Tao1983}, is obtained:
\begin{equation}  
V({\bf s}) = \delta_{s_1+s_2,s_3+s_4}\frac{e^2}{L_y} \int_{-\infty}^\infty  dq  e^{-\frac{\ell^2}{2}q^2+iq\frac{2\pi\ell^2}{L_y}(s_3-s_2)}
 \frac{e^{-\frac{2\pi^2\ell^2}{L_y^2}(s_3-s_1)^2 }}{\sqrt{ q^2 + \frac{4\pi^2}{L_y^2}(s_3-s_1)^2 }}
\end{equation}
A great simplification occurs in the thin torus limit $L_x\ll \ell$, when dependence on $s_3-s_2$ 
disappears, and the matrix element only depends on $|s_3-s_1|$ (modulo $g$). 
In this limit the integral yields a Bessel function:
$$  
V(|s_1-s_3|) = \frac{e^2}{L_y}   e^{-\frac{\pi^2 \ell^2}{L_y^2}(s_1-s_3)^2}K_0 \left (\frac{\pi^2\ell^2}{L_y^2}(s_1-s_3)^2 \right )
$$ 
The Coulomb operator becomes $\frac{1}{2}\sum_{s,t,u} V(|s-u|)a^\dagger_s a^\dagger_t
a_{s+t-u} a_u $. The Fourier transform $c_k = \frac{1}{\sqrt{g}} \sum_{s=1}^{g}\exp(-\frac{2\pi i}{g}ks) a_s$ then 
makes the Hamiltonian diagonal:
\begin{equation}
\label{eq:latticegas}
H=\frac{\hbar\omega}{2}\sum_k n_k + \frac{1}{2}\sum_{k,k'=1}^{g} \tilde V(k-k') n_k n_{k'} 
\end{equation}
where $n_k=c_k^\dagger c_k$ and $ \tilde V(k) = \sum_{s=1}^{g} V(s)\exp(\frac{2\pi i}{g}ks)$. When $g\gg $  range of $\tilde V(k)$,  
the problem of the ground state was exactly solved by Hubbard, 
in the context of classical 1D lattice gases \cite{Hubbard1978}.
He was searching for the sequence $\left\{n_k\right\}=\left\{0,1\right\}^L$ minimising the energy
$\sum_{k,k'} \tilde V(k-k') n_k n_{k'} $
at fixed density $\rho=\sum_k n_k/L$.
Remarkably, Hubbard found that the answer is independent of the form of the interaction,
provided the potential is convex with a finite asymptotic value, namely
$\tilde V(k)\searrow 0$, $\tilde V(k+1)+\tilde V(k-1)>2\tilde V(k)$.
If the irreducible-fraction representation of the density is $\rho=p/q$,
the ground-state sequence is periodic of period $q$, and is such that
the position of the $n$-th particle is $\lfloor n q/p \rfloor$,
where $\lfloor \cdot \rfloor$ is the floor function.
Hubbard's periodic solution gives the vector of periodic occupation numbers of Landau states $\theta_s$. 
In coordinate space, it is a Slater determinant of Jacobi theta functions.
Minimising the Hamiltonian (\ref{eq:latticegas}) is equivalent to finding
which of the Hubbard states is the ground state of the energy
$E= \frac{1}{2}\sum_{k,k'} \tilde V(k-k') n_k n_{k'} -\mu \sum_k n_k$
as a function of the chemical potential $\mu$.
This was solved by Bak and Bruinsma \cite{BakBruinsma},
who showed that the ground-state density $\rho(\mu)$
has a very interesting behaviour: it is a complete Devil's staircase, 
i.e. the Cantor function with plateaux at every rational number.
(Recently, a related fractal hierarchy was discovered in 
the out of equilibrium counterpart of the model studied by Hubbard
\cite{RotondoDLG}).

\begin{figure}[h]
\begin{center}
\includegraphics[width=7cm]{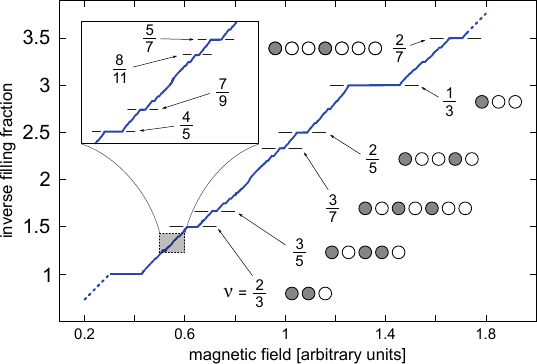}
\caption{The Devil's staircase for the FQHE in the thin torus limit. It shows a correct trend in the asymmetries of widths for particle-antiparticle
exchange.}\label{DevilFQHE}
\end{center}
\end{figure}

While the sequence of zeros and ones is independent of the potential, the widths of the plateaux do depend on it, and are given by the formula \cite{BakBruinsma,BurkovSinai}:
$$  \Delta\mu = 2q \sum_{k=1}^\infty k[\tilde V(kq+1)+\tilde V(kq-1)-2\tilde V(kq)]. $$
To export the phase diagram of the lattice gas $(\mu, \rho)$ to the thin torus FQHE diagram $(B,1/\nu)$ one has to rescale the Hamiltonian to cancel the dependence on $B$ of the interaction term (of course, a global factor does not change
the ground state). Only the rescaled kinetic energy depends on the magnetic field, 
and corresponds to $\mu$ in the lattice gas.
The outcome is a diagram, fig.~\ref{DevilFQHE}, where the plateaux widths qualitatively reproduce
the experimental observations. Notice that, while fractions with the same denominator have equal widths
in $\mu$, this symmetry is lost in the $(B,1/\nu)$ plot.
Another feature modifies the diagram: even-denominators thin torus ground states are forbidden because of Fermi statistics and modular invariance, as recently shown by Seidel \cite{Seidel}.

Although the plot of the Devil's staircase in the thin torus limit is an intriguing result, we know that thin torus (TT) ground states can not describe truly two-dimensional FQH states, which are much better approximated by Laughlin's and composite fermion wavefunctions. However, this result encouraged us to investigate more in detail the algebraic structure of Laughlin' wavefunctions, that recently Haldane and Bernevig found to belong to a class of special orthogonal polynomials, the Jack polynomials. In the following we introduce this fascinating mathematical topic, introducing these polynomials as exact eigenfunctions of a well known integrable quantum system: the Calogero-Sutherland model.

\section{\bf Calogero-Sutherland and Laplace-Beltrami operators} 
Consider $N$ particles on the unit circle, that may cross and interact via a two-particle potential $1/r^2$, where $r$ is their chordal distance \cite{Sutherland}: 
\begin{equation}
H_{CS} (\beta) =  -\sum_{k=1}^N \frac{\partial^2}{\partial\theta_i^2} + \frac{1}{2}\beta(\beta -1)  \sum_{j<k} \frac{1}{\sin^2 \frac{1}{2}(\theta_j-\theta_k)}
\end{equation}
$\beta >0$ is a free parameter. 
The total momentum $P=-i\sum_k \partial/\partial \theta_k$ is conserved. Since
$H_{CS}  = \sum_{k=1}^N A^\dagger_k  A_k  +E_0$ where
 \begin{equation}
   A_k = -i\frac{\partial}{\partial\theta_k} +i \frac{\beta}{2} \sum_{j\neq k} {\rm cotg} \frac{1}{2}(\theta_k-\theta_j),
 \qquad  E_0 = \frac{\beta^2}{12}N(N^2-1),
\end{equation}
the ground state is obtained by solving $A_k \psi_0 =0$ for all $k=1...N$, and has energy $E_0$. 
The extension beyond the sector $\theta_1<\dots <\theta_N$ brings a problem of sign. For bosons or fermions:
$$\psi_0 (\theta_1,\dots ,\theta_N) = \prod_{j>k}   |\sin \frac{1}{2}(\theta_j-\theta_k) |^\beta \times \Big\{ 
\begin{array}{c} 1 \\ {\rm sign} (\theta_j-\theta_k) \end{array} $$
In the variables $z_k=\exp(i\theta_k )$ the Hamiltonian and the total momentum become
\begin{equation}
H_{CS}= \sum_{k=1}^N (z_k \partial_k)^2 -2\beta (\beta-1) \sum_{j<k} \frac{z_jz_k}{(z_j-z_k)^2}, \quad
P=\sum_{k=1}^N z_k \partial_k \label{HCS}
\end{equation}
Note that 
$\prod_{j>k}\sin^2 \frac{1}{2}(\theta_j-\theta_k) = \frac{\Delta(z) \Delta(1/z)}{2^{N(N-1)}} $
where $ \Delta (z)$  is the Vandermonde determinant.
Since $\bar z=1/z$, the  expression coincides with $|\Delta(z)|^2$ up to a constant. The bosonic ground state is 
\begin{equation}
 \psi_0(z)=|\Delta (z)|^\beta 
 \end{equation}
with null total momentum.
The excited states are searched in the factored form $|\Delta (z)|^\beta f(z)$. The eigenvalue equation $\psi_0^{-1}H_{CS}\psi_0 f=E f $ becomes $H_{LB}f =(E-E_0)f$, where the Laplace Beltrami operator is:
\begin{equation}
 H_{LB}=\sum_{k=1}^N (z_k\partial_k)^2 + \frac{\beta}{2} \sum_{j\neq k} \frac{z_k + z_j}{z_k-z_j} (z_k\partial_k - z_j\partial_j)  
 \label{LB}
\end{equation}
The total momentum is $P'f=\psi_0^{-1}P\psi_0f = \sum_k z_k\partial_k f=Pf$ and commutes with $H_{LB}$. 
When promoting $z$ from the unit circle to the whole complex plane, some differences arise. It is 
instructive to summarize the problem. \\
In the Bargmann space of entire functions $f(z)$, $z=(z_1\dots z_N)$, 
such that 
$ \int  \frac{d^{2N}z}{\pi^N} e^{-\sum_k |z_k|^2} |f(z)|^2 <\infty  $, 
we look for a potential $V$ and energy value $E$ such that $[\sum_{k=1}^N (z_k\partial_k)^2 + V(z_1\dots z_N) -E] \Delta(z)^\beta =0 $, i.e. $ E(\beta )-V({\bf z}) =\frac{1}{\Delta^\beta}\sum_{k=1}^N (z_k\partial_k)^2 \Delta^\beta $.
We recover the Hamiltonian $H_{CS}$ of the Calogero-Sutherland model, eq.\eqref{HCS}, 
with eigenvalue $E(\beta) = \frac{1}{6}\beta^2 N(N-1)(2N-1) $.\\
If the excited states are searched with the factored form $\psi ({\bf z}) = \Delta({\bf z})^\beta f({\bf z})$, 
then $f$ is an eigenfunction of the operator
\begin{eqnarray} 
 [H'-E(\beta)]f =&& \left[ \Delta(z)^{-\beta} \sum_{k=1}^N (z_k\partial_k)^2  \Delta(z)^\beta + V(z)-E(\beta)
 \right ] f({\bf z})  
 \nonumber \\
&& = H_{LB} f ({\bf z})+\frac{1}{2}\beta(N-1) P f ({\bf z})
\end{eqnarray}
We obtained the Laplace-Beltrami operator \eqref{LB}, corrected by a boost. In Barg\-mann space, the
operators $H'$ and $H_{LB}$ are not Hermitian, but they gain a nice mathematical property: in the basis of monomials of $N$ variables, they are lower triangular. This allows to find all the 
eigenvectors, the celebrated Jack polynomials, that we introduce in the next section.

\section{\bf Jack polynomials}
The theory of Jack polynomials requires the concepts of partition and squeezing of a partition. They will fit naturally in the Fock space formalism.

Partitions are sequences $\lambda =\{\lambda_1,\dots , \lambda_N \}$ of integers $\lambda_1 \ge \dots \ge \lambda_N \ge 0$. The length $|\lambda |$ of a partition is the sum of its integers. A partition can be specified by the sequence of multiplicities of the integers. Physicists think of them as occupation numbers ($n_0,n_1,\dots $) of particles on a semi-infinite lattice. For instance, $\{4,1,1,0\}$ has 1 zero ($n_0=1$), 2 ones $(n_1=2)$, and 1 four $(n_4=1)$; then it can be written as $(120010\dots )\equiv (12001)$ i.e. four particles in positions 0,1,1 and 4 (see Fig. 4a).\\ 
We define the squeezing of a partition in the picture of occupation numbers. A squeezing produces a new partition
where two particles are moved toward each other. A squeezing of $(12001)$ is $(11110)$ (one of the two particles 
in site 1 and the particle in site 4 are moved to sites 2 and 3).
There are two other squeezes of $(12001)$: $(03010)$ (we move the particle in site 0 and the one in site 4 by one or three sites) and $(02200)$ (we move 0 and 4  by two sites). A pictorial representation of some of these squeezing is 
given in Fig. 4b and 4c.\\
We define a partial order among partitions with same number of particles: $\lambda<\mu$ if the partition $\mu$ can be obtained by one or more squeezings of the partition $\lambda$.

\begin{figure}[t]
	\begin{center}
		\includegraphics[width=8 cm,keepaspectratio]{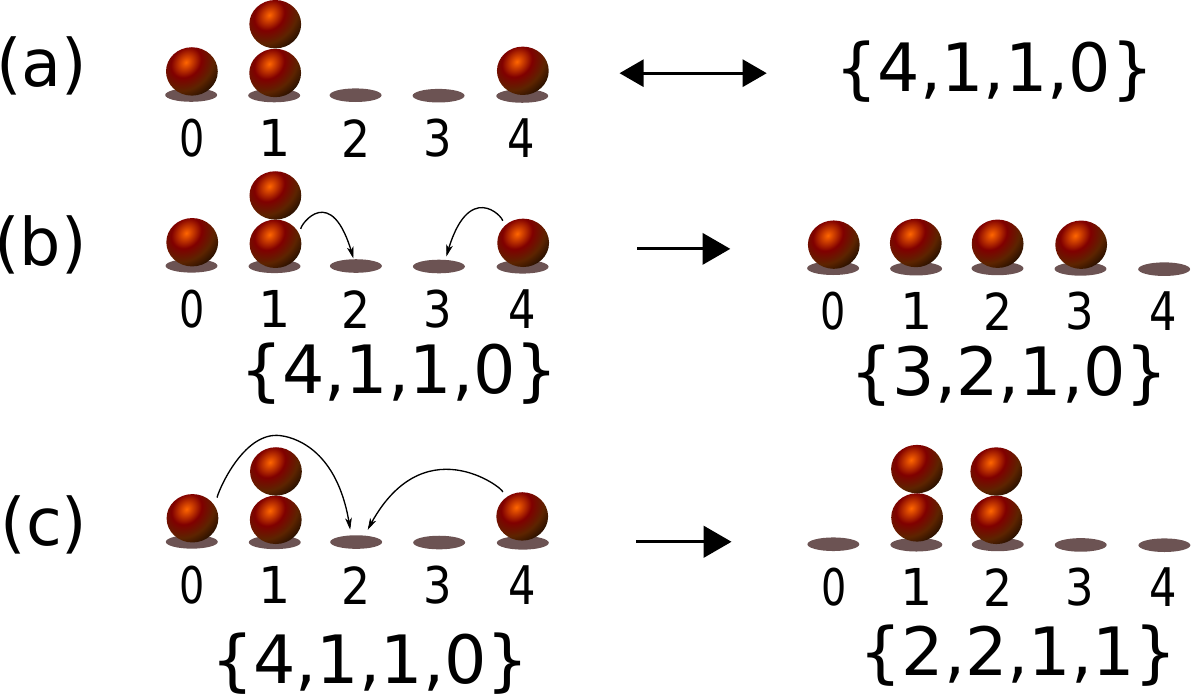}
		\caption{(a) A configuration of occupancies corresponds to a partition. Here the lattice configuration on the left corresponds to the partition $\{4,1,1,0\}$. (b) and (c) Lattice and partition representation of two of the possible squeezings of the partition $\{4,1,1,0\}$.} \label{figsqueeze}
	\end{center}
\end{figure}

A partition $\lambda =\{\lambda_1,\dots,\lambda_N\}$ specifies a monomial ${\sf m}_\lambda ({\bf z})$, 
${\bf z}\in \mathbb C^N$: 
\begin{equation}\label{mon}
{\sf m}_{\lambda}({\bf z}) = \sum_\pi z_{\pi_1}^{\lambda_1} \dots z_{\pi_N}^{\lambda N}
\end{equation}
The sum covers permutations $\pi $ of $1,\dots, N$ without repetition of equal terms. 
 A monomial is a symmetric polynomial of degree $|\lambda |$. For example: ${\sf m}_{3,1,1}({\bf z}) =z_1^3 z_2z_3+ z_2^3z_1z_3 +z_3^3z_1z_2$.
The symmetric monomials ${\sf m}_\lambda (z)$, $z\in \mathbb C^N$, are an orthogonal basis in Bargmann space.\\
\begin{theorem}[Sogo, \cite{Sogo}]
In the basis of monomials with fixed degree, the Laplace-Beltrami operator $H_{LB}$ is triangular, i.e.:
$$ H_{LB} {\sf m}_\lambda ({\bf z}) = E_\lambda {\sf m}_\lambda ({\bf z}) + \sum_{\mu <\lambda } C_{\lambda\mu} {\sf m}_\mu  ({\bf z}) $$
with diagonal values (the eigenvalues of $H_{LB}$):
\begin{equation}
E_\lambda = \sum_{j=1}^N \lambda_j^2  + \beta \sum_{j<k}(\lambda_j -\lambda_k) = 
\sum_{j=1}^N \lambda_j^2 +\beta (N+1-2j) \lambda_j 
\end{equation}
$C_{\lambda\mu}=0$ if $\mu$ is not a partition squeezed from $\lambda $,  
$$C_{\lambda\mu} = 2\beta (\lambda_i-\lambda_j) { {n(\mu_i)}\choose {2}} \quad (\mu_i=\mu_j), \quad 
C_{\lambda\mu} = 2\beta (\lambda_i-\lambda_j) n(\mu_i) n(\mu_j) \quad (\mu_i\neq \mu_j)$$
where the squeezing takes $(\lambda_i,\lambda_j)\to(\mu_i,\mu_j)$ ($i<j$).
\end{theorem}
The diagonalization of the matrix brings the polynomial eigenfunctions of $H_{LB}$ of degree $|\lambda |$: the symmetric Jack polynomials 
$J_\lambda ({\bf z})$.
\begin{eqnarray}
  J_\lambda ({\bf z}) = {\sf m}_\lambda ({\bf z}) + \sum_{\mu<\lambda } b_{\lambda,\mu} {\sf m}_\mu ({\bf z}) 
\end{eqnarray}
Lapointe and Lascaux \cite{Lapointe_Lascaux} gave a determinantal expression for $J_\lambda $ (with $b_{\lambda\lambda}\neq 1$).

\subsubsection*{\bf Example} $N=3$, $|\lambda | =3$.\\
The partitions with $N=3$ particles and length 3 are $\{1,1,1\}$, $\{2,1,0\}$, $\{3,0,0\}$. Then, the matrix $H_{LB}$ is $3\times 3$. The basis of monomials are:
${\sf m}_{111} = z_1 z_2 z_3$, ${\sf m}_{210} = z_1^2 z_2 + z_1^2z_3 + z_2^2 z_1 + z_2^2 z_3 + z_3^2z_1 + z_3^2z_2$, ${\sf m}_{300} = z_1^3 +z_2^3 + z_3^3$.\\
\begin{equation}
H_{LB}\left[ \begin{array}{c} m_{111}\\m_{210}\\m_{300} \end{array}\right] = 
\left[ \begin{array}{ccc} 3 & 0 & 0  \\
12\beta & 5+4\beta & 0  \\
0 & 6\beta & 9+6\beta  
\end{array}\right] \left[ \begin{array}{c} m_{111}\\m_{210}\\m_{300} \end{array}\right]  \nonumber
\end{equation}
Diagonalization brings out $3$ Jack polynomials: $J_{111} = {\sf m}_{111}$, $
J_{210}= {\sf m}_{210} + \frac{6\beta}{1+2\beta} {\sf m}_{111} $, and $ J_{300}= {\sf m}_{300} + \frac{3\beta}{2+\beta} {\sf m}_{210} + \frac{6\beta^2}{(1+\beta)(2+\beta)}{\sf m}_{111} $
with eigenvalues $3$, $5+4\beta$, $9+6\beta$.

\section{\bf Fock space}
Let us introduce the Fock space formalism for the lowest Landau level.\\
If $\ket{\lambda}$, $\lambda= 0, 1, 2, \dots $ is an orthogonal basis for the single-particle Hilbert space, the 
normalized basis is  $\tfrac{1}{\sqrt{\nu_\lambda}} \ket{\lambda}$, with $\nu_\lambda = \braket{\lambda | \lambda}$.\\
For the lowest Landau level: $\langle z|\lambda\rangle = z^\lambda $, $\nu_\lambda =\lambda !$,
$\lambda =0,1,\dots $ \\
The orthogonal bases for the Hilbert space of $N$ bosons or fermions are given respectively by:
\begin{equation}\label{bases}
\begin{split}
& \ket{\textrm{m}_{{\lambda}}} = \sum_{\pi} \ket{\lambda_{\pi_1} \dots  \lambda_{\pi_N}}, \\
& \ket{\textrm{sl}_{{\lambda}}} = \sum_{\pi} (-)^{\pi} \ket{\lambda_{\pi_1} \dots  \lambda_{\pi_N}}, 
\end{split}
\end{equation}
where $\lambda = \{\lambda_1, \dots, \lambda_N\}$, with $\lambda_1 \geq \dots \geq \lambda_N$, is a partition which specifies the single particle states (for fermions equality is forbidden). The sum runs over permutations of indices $\{1, \dots, N\}$ without repetition of equal terms and $(-)^{\pi}$ is the parity of the permutation.
For example: 
\begin{equation}\label{ex}
\begin{split}
& \ket{\textrm{m}_{440}} = \ket{440} + \ket{404} + \ket{044}, \\
& \ket{\textrm{sl}_{410}} = \ket{410} - \ket{401} + \ket{041} - \ket{014} + \ket{104} - \ket{140}.
\end{split}
\end{equation}
We recover the many-particle wavefunctions of lowest Landau level, in Bargmann space:
$\braket{\vec{z}|\textrm{m}_{{\lambda}}}   = \textrm{m}_{{\lambda}}(\vec{z}) $ (the monomials in Eq.\eqref{mon}) and $\braket{\vec{z}|\textrm{sl}_{{\lambda}}}   = \textrm{sl}_{{\lambda}}(\vec{z})$
(Slater determinants).\\
The bases \eqref{bases} have the equivalent notations:\\
-  \mbox{$\ket{\lambda_1, \dots, \lambda_N}$}, specifying the single-particle quantum numbers ($\lambda_1 \geq \dots \geq \lambda_N $); \\
- $\ket{n_0, \dots, n_{\infty}}$, specifying how many particles share the same quantum number.\\
We introduce the canonical algebra of creation and destruction operators of 1-particle states $\frac{1}{\sqrt{\nu_\lambda}} \ket{\lambda}$, for bosons (+)  and fermions (-):
\begin{equation}
\begin{split}
a_\lambda^\dagger \ket{\dots n_\lambda \dots} &= (\pm 1)^{n_0 + \dots + n_{\lambda-1}} \sqrt{n_\lambda + 1} \ket{\dots n_\lambda+1 \dots}, \\
a_\lambda \ket{\dots n_\lambda \dots} &= (\pm 1)^{n_0 + \dots + n_{\lambda-1}} \sqrt{n_\lambda} \ket{\dots n_\lambda-1 \dots},\\ 
a_\lambda \ket{\dots n_\lambda \dots} &= 0 \quad \text{if } n_\lambda = 0. \\
\end{split}
\end{equation}
The operators allow for an efficient description of the squeezing operation introduced before, which is implemented by the following operator:
\begin{equation}\label{sqop}
s_{u, m, k} = \sqrt{\frac{\nu_{u-k} \nu_{m+k}}{\nu_u \nu_m}} a^\dagger_{u+k} a^\dagger_{m-k} a_m a_u 
\end{equation}
for $0 \leq u < m$ and $0 < k < m-u$.
Its action on the basis $\ket{\textrm{m}_{{\lambda}}}$  is:
\begin{equation}\label{sonmon}
s_{u, m, k} \ket{\textrm{m}_{{\lambda}}} = 
\begin{cases}
0 & \!\!\!\!\!\!\!\!\! \text{if $u\notin {{\lambda}}$ or $m\notin {{\lambda}}$}\\
(n_{u+k}+2)(n_{u+k}+1) \ket{\textrm{m}_{{{\mu}}}}& \; k = \frac{m-u}{2} \in \mathbb{N}\\
(n_{u+k}+1)(n_{m-k}+1) \ket{\textrm{m}_{{{\mu}}}}& \; \text{otherwise},
\end{cases}
\end{equation}
where ${{\mu}}$ is obtained from ${{\lambda}}$ by substituting two particles of quantum numbers $u$ and $m$ with two particles of quantum numbers $u+k$ and $m-k$, hence by ``squeezing two particles by $k$''.  The action on $\ket{\textrm{sl}_{{\lambda}}}$ is
\begin{equation}
\begin{split}
s_{u, m, k} \ket{\textrm{sl}_{{\lambda}}} 
= (-)^{N_\mathrm{sw}}\ket{\textrm{sl}_{{\mu}}},
\end{split}
\end{equation}
where $N_\mathrm{sw}$ is the number of exchanges restoring the decreasing order of the sequence.

\section{A map for FQHE effective states} 
In analogy with Sogo's theorem, establishing a correspondence between monomials and Jack polynomials, we obtain an operator that generates symmetric or antisymmetric eigenstates of the Laplace-Beltrami operator, from simple root states. Some eigenstates are effective ground states of the FQHE.
 
Let $H_0$ be a Hermitian operator, expressed solely in terms of number operators. Its eigenstates are the vectors 
$|{\sf m}_\lambda \rangle$ (bosons) or $|{\sf s}_\lambda \rangle$ (fermions). We collectively denote them as $|\lambda \rangle $. Let $H=H_0+V$, where $V$ is a linear combination of squeezing operators. 
Note that if $|\lambda \rangle $ is an eigenstate of $H_0$ with a finite number of
particles, then there is $n$ such that $V^n$ annihilates $|\lambda \rangle $.

\begin{theorem}\label{teo2}
If $E_\lambda $ is a non degenerate eigenvalue of $H_0$, with eigenvector $|\lambda\rangle $
then $E_\lambda $ is an eigenvalue of $H$ with eigenvector
\begin{equation}
|\psi_\lambda \rangle =  \sum_{k=0}^n [(E_\lambda -H_0)^{-1} V]^k |\lambda\rangle  \label{geometric}
\end{equation}
\begin{proof} Multiplication by $(E_\lambda -H_0)^{-1}V$ gives the identity
\begin{equation}
(E_\lambda -H_0)^{-1}V |\psi_\lambda \rangle = |\psi_\lambda \rangle - |\lambda \rangle  \label{eq_for_psi}
\end{equation}
Multiplication by $(E_\lambda -H_0)$ gives $V|\psi_\lambda \rangle = (E_\lambda -H_0)|\psi_\lambda \rangle$. 
\end{proof}
\end{theorem}
The reverse is true: if $|\psi\rangle $ is an eigenstate of $H$ with non-degenerate eigenvalue $E$, then 
$|\lambda \rangle = |\psi\rangle -(E-H_0)^{-1} V |\psi \rangle $ is an eigenvector of $H_0$ with same eigenvalue.

By construction $|\psi_\lambda \rangle $ is a linear combination of basis states that are squeezed from $|\lambda \rangle $:
$ |\psi_\lambda \rangle = \sum_{\mu\le \lambda} b_{\lambda\mu} |\mu\rangle  $.
Eq.\eqref{eq_for_psi} gives $b_{\lambda\lambda}=1$ and the recursive relation:
\begin{equation}
b_{\lambda\mu} = \frac{1}{E_\lambda -E_\mu}\sum_{\mu<\mu'<\lambda } \langle \mu|V|\mu'\rangle  b_{\lambda,\mu'} \quad (\mu\neq\lambda ) \label{eq_recursion}
\end{equation}
The sum only involves states that yield $|\mu \rangle $ after a single squeezing contained in $V$ and that descend from the root $|\lambda \rangle $ by one or
more squeezings.

\begin{figure}[h!]
 	\begin{center}
 		\includegraphics[width=8 cm,keepaspectratio]{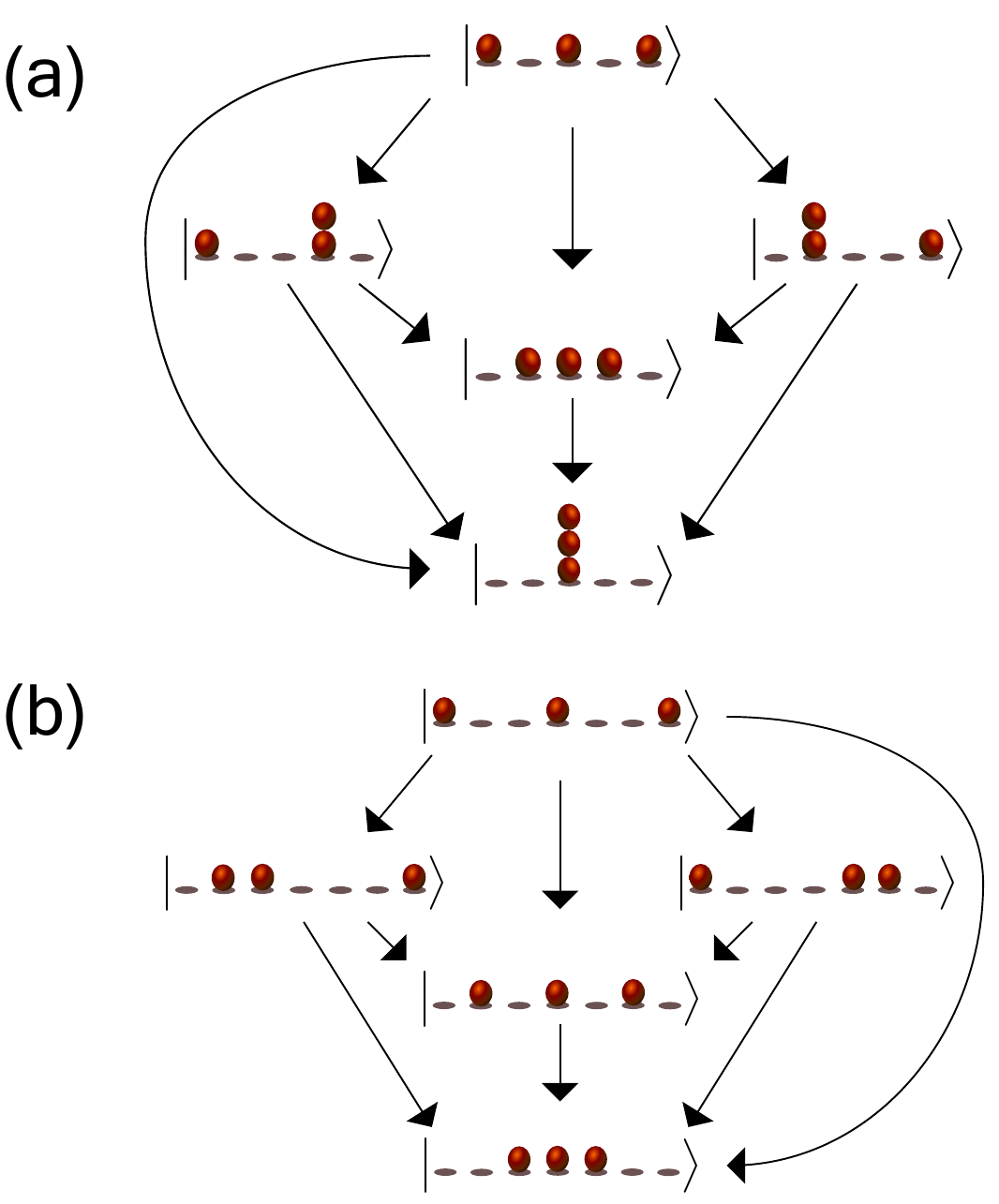}
 		\caption{The repeated action of the squeezing operator in Eq. \eqref{squeezing_op} on a bosonic (a) and a fermionic (b) state with three particles. Each state is specified by the occupancy numbers, and each arrow represents the action of a squeeze. The operator in Eq. \eqref{squeezing_op} automatically takes into account the possibility of multiple occupancies of the bosonic case.}\label{fig:squeeze_chain}
 	\end{center}
\end{figure}

We can use this theorem to obtain the FQHE effective states that are a Jack polynomial times a Vandermonde determinant, since such wavefunctions 
are eigenstates of the ``Generalized Laplace Beltrami'' operator:
\begin{equation}\label{}
H=\sum_{k=1}^N (z_k\partial_k)^2  + K \sum_{j\neq k} \frac{z_k+ z_j}{z_k-z_j} \left (z_k\partial_k - z_j\partial_j\right)
- K\sum_{j\neq k} \frac{z_j^2+z_k^2}{(z_j-z_k)^2}(1-\pi_{jk})
\end{equation}
where $\pi_{jk}$ is the exchange operator of particles $j,k$.  In the bosonic sector $\pi_{jk}=1$, and we recover the Laplace Beltrami operator Eq.\eqref{LB}. $K=  (1-2q\pm 1)/(p+1) $ ($+1$ for bosons, $-1$ for fermions), with filling fraction $\nu=p/q$.\\
By evaluating $H$ in second quantization we obtained $H=D + K S$, where $S$ is a linear combination of squeezing operators
\begin{equation}\label{squeezing_op}
S=\sum_{s=0}^{\infty} \sum_{t=1}^{\infty} \sum_{u=1}^{\infty} u \sqrt{\frac{(s+t)! (s+u)!}{(s+t+u)! s!}} \: a^\dagger_{s+t} \: a^\dagger_{s+u} \: a_{s+t+u} \: a_{s}
\end{equation}
and $D$ incorporates the kinetic term and the diagonal part of the two-body potential (up to an irrelevant additive constant for fermions):
\begin{align}\label{D_op}
D = \sum_{m=0}^\infty m^2 a^\dagger_m a_m + \frac{K}{2} \sum_{m'<m} (m-m') a^\dagger_m a_m a^\dagger_{m'} a_{m'}.
\end{align} 
Remarkably, we obtain Laughlin's wavefunction at filling fraction $1/q$ if we use as root the thin torus 
ground state with same filling fraction. In Fig. \ref{fig:squeeze_chain} we show the action of the squeezing operator on the thin torus ground state for $N=3$ and $\nu = 1/3$, generating Laughlin's wavefunction $q=3$.
 
\section{Conclusions}
First, we showed that the FQHE Hamiltonian can be mapped,
in the thin torus limit, to a one-dimensional (classical) lattice gas model,
whose main ingredient is a long-range repulsive interaction.
The map brings one to interpret the ground states as Hubbard states
and to prove their incompressibility, i.e., the fact that they are stable in finite
intervals of magnetic field.
Importantly, this enables to produce a Devil's staircase phase diagram for the filling fractions,
which shows realistic features, such as the asymmetry of the plateau widths,
qualitatively in accordance with the experimental landscape.

Second, we obtained an explicit Fock-space representation of FQHE wavefunctions. The generic recursive relation \eqref{eq_recursion} recovered the known relations for the coefficients of the expansion
of Jack polynomials \cite{Thomale}.\\
To address the question {\em what are Jacks doing on the Devil's staircase?}, it would be useful to identify in the
open plane and in the symmetric gauge, a mechanism (analogous to the thin torus mapping) that selects the
monomials or Slaters that are mapped by the operator \eqref{geometric}, specified by 
\eqref{squeezing_op} and \eqref{D_op}, to the Laughlin state or others
relevant for FQHE. Another intriguing question is to understand whether the composite fermion wavefunctions that generalise the Laughlin's ansatz at more complicated filling fractions, are related in some way to the theory of Jacks. This is already the case for Moore-Read and Read-Rezayi wavefunctions, and would be one more key indication about the fundamental role that Jack polynomials play in the theory of FQHE.

\end{document}